# Development of an integrated device based on the gain/phase detector and Arduino platform for measuring magnetoelastic resonance

Wenderson R. F. Silva[*], Rafael O. R. R. Cunha, Joaquim B. S. Mendes[†]

Departamento de Física, Universidade Federal de Viçosa, 36570-900 Viçosa, Minas Gerais, Brazil.

**Abstract**

This study presents the development of an integrated device for measuring magnetoelastic resonance, utilizing a gain/phase detector and the Arduino platform. The device stands out for its simplicity, low cost, and efficiency. It employs the Arduino's ATmega328P microcontroller, coupled with the AD8302, and an intuitive graphical interface developed in Python, which facilitates the measurement of magnetoelastic signals in sensors. The device's validation is carried out through measurements of mass deposits on magnetoelastic sensors, confirming its accuracy and functionality. This advancement offers a practical and portable solution for detecting magnetoelastic resonances, promoting the use of these technologies in field environments and in diverse scientific applications. The integration of components into a compact, robust, and low-cost electronic circuit, along with an easy-to-operate graphical interface, significantly simplifies the complexity of the components, making the equipment complete and readily operational.

Corresponding authors: [*] wenderson.f@ufv.br, [†] joaquim.mendes@ufv.br

## 1. Introduction

Magnetoelastic materials (ME) stand out for their notable potential in the development of sensors that play a crucial role in the scenario of scientific research related to the detection of physical, chemical and biological agents, as evidenced by recent studies [1]. Sensors manufactured with these materials operate based on the phenomenon of magnetostriction, in which the magnetic response of the material is directly influenced by mechanical deformations. The ability of these materials to convert changes in shape, resulting from applied voltages, into measurable variations in their magnetic state, gives these sensors high sensitivity and precision. The main advantages of which are: i) the possibility of wireless detection, ii) high sensitivity [2-4], iii) chemical and physical resistance [5-6]. Furthermore, it is a versatile and easily viable technique for its production and detection, dispensing the need for electrical contacts and lithographic processes. Added to this, the remote query nature of the ME sensor platform can offer a major advantage when a direct probe or electrical contact with the sensor is not a viable alternative.

Given the growing importance of detection platforms based on ME materials, the need to develop instrumentation for the detection of ME signals becomes evident. Laboratory devices such as vector network analyzers (VNA) [3], real-time spectrum analyzers (RTSA) [7], impedance analyzers [8], Lock-in amplifier application circuits [9] and oscilloscopes [10] are some of the instruments that can be used as detectors of these signals. However, the applicability of these instruments in field environments is limited due to their weight, high cost and complexity. Therefore, the search for more practical and accessible solutions for implementing this technology is essential, aiming for its effective implementation in real-world conditions. Several studies have been dedicated to this issue [11-14]. Recently Sang et al. [15] proposed a portable microcontroller-based system, integrated through System on Chip (SoC) architecture, to sample resonance frequency changes in an ME sensor used in the detection of human serum albumin (HSA). Shen et al. [16] and Zeng et al. [14] proposed time-domain instrumentation to determine the resonance frequency in ME sensors. Yang et al. [17] used a portable system for measurements in aquatic environments.

However, these devices feature relatively complex electronic circuits, requiring significant infrastructure for manufacturing, making them impractical in scenarios where cost efficiency and ease of production are priorities. On the other hand, conventional laboratory equipment can require a complex and high-cost experimental setup, involving various equipment, specific cables and automation knowledge for integration, making detection systems expensive and requiring specialized training. In this scenario, we propose a device with notable advantages, such as low cost and electronic optimization, to simplify the complexity of the components. Using the Arduino ATmega328P microcontroller and an easy-to-operate graphical interface developed in Python, the device not only stands out for its effectiveness, but also contributes to making magnetoelastic technologies more accessible, enabling their application in various areas of research. The device is based on a compact, optimized, robust and low-cost electronic circuit for detecting ME signals in sensors, with the potential to be used in field environments. The complete integration of components into an intuitive graphical user interface (GUI), also developed in this work, makes the equipment complete and readily operational. As a way of validating the device, we carried out measurements of mass deposits on the ME sensor, in addition to measurements in aqueous media and obtained the curve for the delta E effect.

## 2. Implementation of the electronic circuit and graphical user interface (GUI)

The methodology adopted to perform resonance frequency measurements is a frequency sweep. Arduino's ATmega328P microcontroller performs integral control of the electronic system. The AD9833 generates a sinusoidal signal to sweep the excitation coil, which initially passes through a preamp followed by a power amplifier. The excitation circuit couples a DC signal (induction) to the AC sinusoidal signal (modulation), through a power amplifier. The DC signal is necessary to generate a magnetic induction field, so that the ME sensor oscillates with greater amplitude in its anisotropy field. A pickup coil is monitored by a phase and gain detector, which receives a reference signal from the signal generator. Thus, the values associated with the magnitude and phase of the signal are obtained by the pickup coil. When inserting the sensor inside, it enters resonance when the frequency of the excitation signal corresponds to its magnetoelastic resonance frequency, identified as a descending amplitude peak in the signal magnitude spectrum. The entire circuit is powered by a symmetrical 9 V supply. The Arduino platform exchanges data and is powered with +5 V through the computer's USB port. An electronic schematic of the main components of the developed device are detailed in Fig. 1.

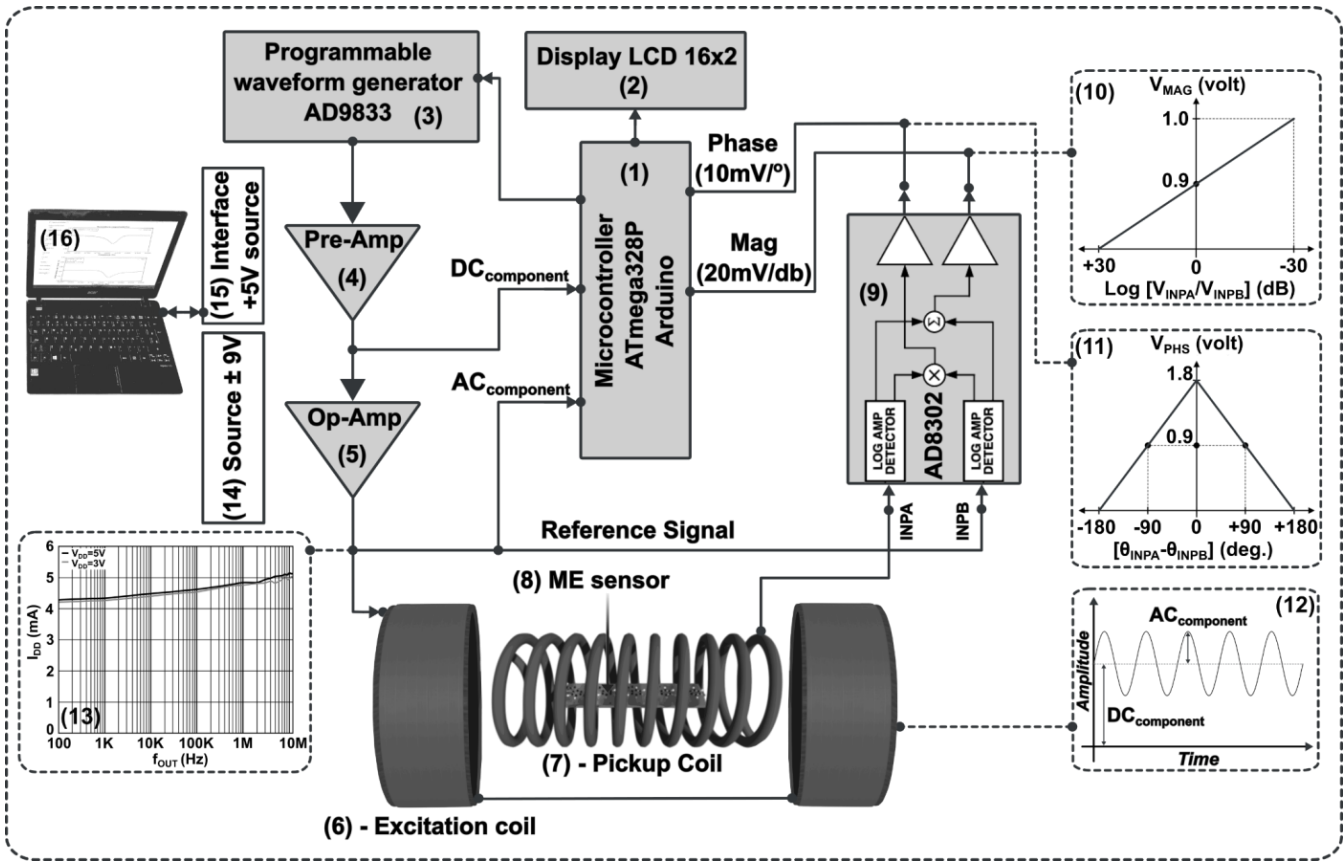

**Figure 1 -** Electronic diagram with the main components of the developed device. (1) The ATmega328P microcontroller controls the signal from the AD9833 (3), reads the signal from the AD8302 (9) and displays the scan parameters on the LCD display (2). Interaction is enhanced through the Graphical User Interface (GUI), allowing adjustments to scan intervals and DC and AC field strengths. Graph (13) shows the linear response to the signal generated by the AD9833. In (10) there is the signal output generated by the AD8302 as a function of the logarithm of the relationship between the signals, $V_{INPA}$ and $V_{INPB}$, in (11), the signal output related to the phase difference ($\theta_{INPA} - \theta_{INPB}$) and, in (12), the coupled signal (DC + AC) delivered to the excitation coil.

**ATmega328P Microcontroller:** the ATmega328P (1) present on the Arduino board is an 8-bit microcontroller with RISC architecture and a clock speed of up to 20 MHz. With integrated peripherals, such as 10-bit resolution ADC and UART serial communication, it offers robust features for control and measurement. Its accessible programming via the Arduino platform makes it ideal for instrumentation projects, being able to easily interact with external sensors and devices. The Arduino is used to establish communication between the excitation (3, 4, 5 and 6) and detection (7 and 9) circuits, as well as carrying out serial communication with the microcomputer. The ATmega328P also receives a 1.8 V signal from the AD8302 (15) for ADC reference, resulting in a resolution of 1.76 mV. The microcontroller is programmed in C++ language through the Arduino IDE, to communicate with the peripherals and the developed graphical interface.

**AD9833 Programmable Waveform Generator:** The AD9833 (2) is a low power programmable waveform generator (12.65 mW power consumption at 3 V), capable of producing sine, triangle and square wave outputs. The output frequency and phase are programmable. The frequency registers have a width of 28 bits, allowing a resolution of 0.1 Hz to be achieved with a clock frequency of 25 MHz, and a frequency range of 0 to 12.5 MHz [18]. The device is programmed to generate sine waves, which are produced with an amplitude of 0.6 V and undergo subsequent amplification processes.

**Pre and power amplifier:** the signal from the AD9833 does not have the ability to excite the coil to stimulate sensor oscillation. Therefore, it is necessary to amplify this signal. A preamplifier (4), based on the LM358 operational amplifier, is used to increase the amplitude of the signal generated by the AD9833 from 0.6V to 1.2V. The signal then goes to a second amplification stage, using the OPA544 operational power amplifier (5). At this stage, it becomes possible to adjust the amplitude of the AC signal from 1.2 V to 8.0 V, with considerable current gain. A DC offset signal (necessary to generate the induction magnetic field) is added to the AC signal, allowing adjustments in the range of -5 V to +5 V. The resulting new AC signal is then filtered and sent to the detection system for reference. The AD8302 (only the AC component follows for reference), which is also injected into the excitation coil, generates the induction (DC component) and modulation (AC component) magnetic fields, responsible for exciting the sensor.

**Excitation and pickup coils:** after amplification of the signal, it is applied to the excitation coil (6) (600 turns, 23 mm internal diameter and 50 mm long, with 0.40 mm thick copper wire and resistance of 10.7 Ω), which produces the magnetic field ($H_{DC} + H_{AC}$) necessary to generate the oscillations in the ME sensor. A double-layer pickup coil (7) (200 turns, 4 mm inner diameter and 22 mm long, with 0.18 mm thick copper wire and 4.5 Ω resistance) is allocated concentric to the excitation coil, to which is monitored by the detection circuit described below. A calibration was performed to obtain the DC induction magnetic field from the signals read by the ADS on port $A_1$.

**Phase and gain detector (GPD):** The GPD (9) integrate the device's detection circuit, based on the AD8302 chip from Analog Devices, which measures the magnitude ratio (gain or loss) and the phase difference between two signals [19]. The AD8302 circuit (9) operates as a single-chip vector signal analyzer, where two monolithically integrated identical logarithmic amplifiers play main roles. Such amplifiers perform a logarithmic compression of the input signals, $V_{INPA}$ and $V_{INPB}$, which can vary from -60 dBm to 0 dBm, generating two voltage signals at the output, $V_{MAG}$ and $V_{PHS}$, with a linear range of 0 to 1.8 V. The component also incorporates a subtractor to determine the difference between the input signals. This makes it possible to accurately measure the amplitude and phase of the signal. The operating range extends to 2.7 GHz, with magnitude ratio measurement in a range of ±30 dB, at a sensitivity of $V_{SLP}$ = 30 mV/dB (600 mV/decade), and the phase difference measurement range between the input signals is from 0° to 180°, at a sensitivity of $V_\emptyset$ = 10 mV/°, which further expands the usefulness of this device in various applications that require detailed analysis of radiofrequency (RF) signals. With the AD8302 operating in "Measurement Mode", the expression for the $V_{PHS}$ output phase and the $V_{MAG}$ magnitude gain are given by equations (1) and (2), respectively [19-20]:

$$V_{MAG} = V_{SLP} \log\left(\frac{V_{INPA}}{V_{INPB}}\right) + V_{CP}, \qquad (1)$$

$$V_{PHS} = \pm V_\emptyset(\emptyset(°) - 90°) + V_{CP}, \qquad (2)$$

where $\emptyset$ (°) = $\emptyset_{INPA}$ - $\emptyset_{INPB}$ is the phase difference between the input signals. For $V_{MAG}$, the midpoint is $V_{CP}$ = 900 mV for 0 dB gain, and a range of –30 dB to +30 dB covers the full swing from 0 V to 1.8 V, as shown in the graph (10) in Fig. 1. For the $V_{PHS}$ phase function, the center point is $V_{CP}$ = 900 mV for 90°, and a range from 0° to 180° covers the full swing from 1.8 V to 0 V, as shown in Fig. 1 (11). Equations (1) and (2) can be rewritten to express the magnitude Mag (dB) and the phase difference $\emptyset$ (°) [21-22]:

$$Mag(db) = \log\left(\frac{V_{INPA}}{V_{INPB}}\right) = \frac{V_{MAG}(mV) - 900mV}{30mV}. \qquad (3)$$

$$\emptyset(°) = \emptyset_{INPA} - \emptyset_{INPB} = \pm\left(\frac{V_{PHS}(mV) - 900\ mV}{10\ mV/°} + 90°\right). \qquad (4)$$

**Symmetrical power supply:** To provide symmetrical power supply to the amplification circuit, a symmetrical power supply (1) was designed based on LM7809 and LM7909 voltage regulators. These regulators generate a fixed voltage value of ± 9 V, ensuring a complete wave cycle in the sinusoidal signal. The electronic circuit design, integrating all circuit components, is shown in Fig. 2.

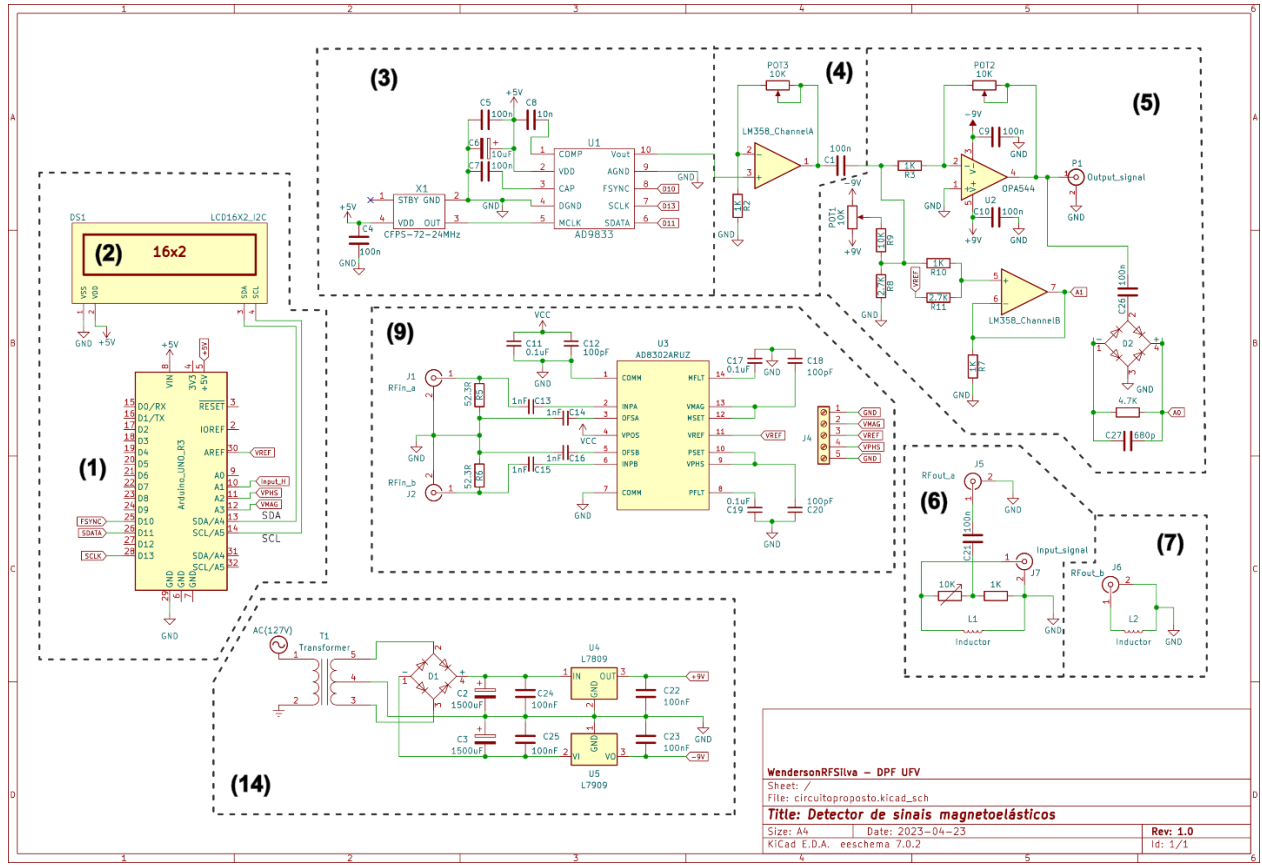

**Figure 2** - Electronic circuit of the device. In (1), ATmega328P microcontroller and (2) LCD display, (3, 4 and 5) AC and DC signal generation circuit. At this stage, the microcontroller's analog terminations $A_0$ and $A_1$ read the AC and DC fields, respectively. Excitation (6) and pickup (7) coils, coil signal reading circuit (9). At this stage, the microcontroller terminations $A_2$ and $A_3$ read the phase and magnitude of the signal read from the pickup coil. Finally, in (14) power supply.

*2.1 Device assembly*

The final device was built by placing the entire electronic circuit in plastic cases. The numbers in Fig. 2 correspond to the component numbers described in Fig. 1. The LCD display (2) was fixed in one of the cases, measuring 12 x 8 x 4 centimeters, together with the excitation and detection circuit, which has its reference, excitation and capture signal outputs connected through SMA terminals. In a second case, measuring 7 x 4 x 3 centimeters, the excitation coil (6) and the pickup coil (7), were fixed, both also equipped with SMA terminals. Communication between the electronic control unit and the coils is established using coaxial cables with a characteristic impedance of 50 Ω. The symmetrical source circuit (14) was placed in a third plastic case, measuring 9 x 4 x 4 centimeters, which communicates with the electronic control unit for power supply via banana-terminated cables. This organization in cases provides a resistant and robust structure for the device, facilitating assembly, maintenance and adaptation as necessary.

*2.2 Developed graphical user interface (GUI)*

The program, developed in the open-source programming language Python, makes the equipment intuitive and easy to operate. For its production, the Tkinter libraries were used for graphic creation, Matplotlib for displaying graphics and Pyautogui for capturing screens. Additionally, libraries such as Serial, Numpy and Scipy are imported for data manipulation and serial communication. The program includes fields for entering parameters, such as serial port address, start and end frequency, sweep step, amplitude and start and end phase. During the scan, data is read from the serial port, processed and displayed in real time on graphs (amplitude and phase) to accompany the scan, where it is also possible to monitor values associated with the DC magnetic field, AC signal amplitude and quality factor of the Q resonance, in addition to saving all data in a .txt file and having the option of exporting images of the graphs in .png format. The image of the main screen of the developed GUI is shown in Fig. 3. A link with the executable version of the graphical interface, together with the Arduino programming code, is available at https://github.com/wphysics/WaveMagScan.

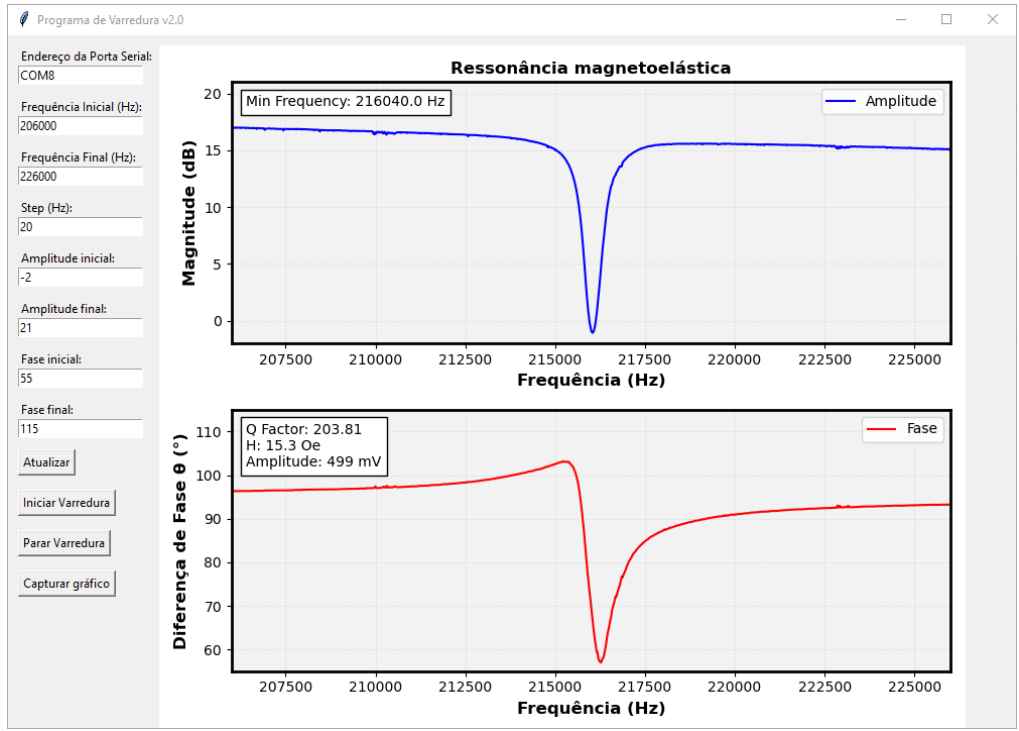

**Figure 3** - Overview of the graphical user interface (GUI) developed for communication with the developed device, through which you can control the parameters associated with the frequency sweep, monitor the sweep through the magnitude phase difference graph, as well as view the values induction fields (DC), modulation (AC) and resonance quality factor. A file in txt format with all these parameters is automatically saved with each scan.

### 3. ME sensor preparation and measurement

Three ME sensors were produced from ribbons of size 11 mm × 2 mm × 28 μm of METGLAS® 2826MB3 ($Fe_{40}Ni_{38}Mo_4B_{18}$) alloy. The sensors were cleaned in ultrasonic baths, first in acetone and then in ethanol, for 10 minutes in each step. Subsequently, a Platinum (Pt) film was deposited on the polished side of five sensors, with thicknesses of 50, 100, 150, 200 and 250 nm, respectively. The resonance frequencies of each ME sensor were measured using the proposed equipment, with a frequency sweep step of 10 Hz. To measure the magnetoelastic resonance frequency, the sensor is inserted inside the pickup coil. Then, through the user interface, the measurements are initialized. All measurements were performed at room temperature (25 °C). For comparison purposes, resonance frequency measurements were performed using a vector network analyzer (R&S®ZNLE18, Rohde & Schwarz, Munich, Germany), operated in $S_{11}$ mode connected to an internal solenoid coil (the detection coil), which generates the AC excitation signal with a step of 5 Hz, to perform a frequency sweep and monitor the reflected signal. The DC magnetic induction field was applied through the external coil (the excitation coil) using a Keysight U8031A DC source.

*3.1. ME resonance frequency detection principle and measurements*

Due to a strong coupling between mechanical and magnetic energies, ME materials can change their dimensions in the presence of a magnetic field. The reflections of which can be studied by evaluating the modulus of elasticity $E$ of the ME material, which will be subject to the "delta $E$" effect, where the modulus of elasticity is a function of the applied magnetic field $H$, $E = E(H)$, given by equation (5) [16, 23-24]:

$$E(H, \sigma) = \left( \frac{1}{E_H} + \frac{9\lambda_s^2 H^2}{\mu_0 M_s H_{A\sigma}^3} \right)^{-1}, \qquad (5)$$

where $E_H$ is the modulus of elasticity without the effect of the magnetic field ($H = 0$), $\mu_0$ is the magnetic permeability of the vacuum, $\lambda_s$ is the magnetostriction constant, $M_s$ is the saturation magnetization and $H_{A\sigma}$ is the magnetoelastic anisotropy field, where the sensor reaches the highest oscillation amplitudes, which depends on the voltage $\sigma$ applied to the material.

A plane wave type disturbance is performed on the ME sensor through the application of AC (modulation) and DC (induction) magnetic fields, $H = H_{DC} + H_{AC}$. By ensuring that the magnetic fields are uniform along the sample (x axis), a one-dimensional ME wave will propagate through the material, which can be described by equation (6):

$$\frac{\partial^2 u_x}{\partial t^2} = \frac{E}{\rho_r (1 - \nu^2)} \frac{\partial^2 u_x}{\partial x^2}, \qquad (6)$$

whose solution is of the type [25]:

$$u_x(t) = u_0 \cos\left(\frac{n\pi x}{L}\right) e^{-i\omega t}, \qquad (7)$$

where $u_0$ is a constant and $\omega = 2\pi f$ is the angular frequency of oscillation of the applied AC field, $n$ is the resonance mode (evaluated at $n = 1$), $L$ is the length of the density sensor. From equation (6) and (7), the ME resonance frequency $f_r$ can be located by equation (8) [1, 25]:

$$f_r = \frac{n}{2L}\sqrt{\frac{E}{\rho(1-v^2)}}, \quad (8)$$

where $\rho$ is the density and $v$ is the Poisson's ratio of the ME sensor. Considering a layer of mass sufficiently thin in relation to the wavelength of the ME resonator, rigidly deposited and uniformly distributed over its surface, which oscillates synchronously with the resonator, so that the mass layer can be treated as part of the resonator. Under these conditions, the average density of the resonator charged with mass $\rho_c$ will be:

$$\rho_c = \rho_r + \frac{\rho_m h}{d}, \quad (9)$$

where $\rho_m$ and $h$ are the density and thickness of the deposited mass layer and $d$ is the thickness of the ME resonator. Solving equation (6) for ($\rho_r = \rho_c$), a relationship between the perturbed $\omega$ and the free $\omega_0$ resonance frequencies by the deposited mass is obtained, given by:

$$\left(\frac{\omega}{\omega_0}\right)^2 = \left(1 + \frac{\rho_m h}{\rho_r d}\right)^{-1}. \quad (10)$$

For a thin layer of dough ($\rho_m h \ll \rho_r d$), equation (10) can be approximated by a series expansion in a linear relationship between frequency change $\Delta f$ and the property $\rho_m h$ of the dough layer [26]:

$$\Delta f_{mass} = -\frac{\rho_m h}{2\rho_r d} f_0, \quad (11)$$

where $\Delta f_{mass}$ is the displacement of the natural frequency $f_0$ of the ME resonator, resulting from the deposition of a thin film on its surface. Assuming that the increase in mass $\Delta m = \rho_m h A$ occurs over the entire surface area $A$ of the ME resonator, equation (11) takes the form [1]:

$$\Delta f = -\frac{\Delta m}{2m_0} f_0, \quad (12)$$

where $m_0$ is the mass of the bare ME resonator. Finally, the mass sensitivity is defined in equation (13). Higher resonance frequencies, which can be achieved with sensors of shorter length $L$, or sensors of lower mass $m_0$ result in higher sensitivity values.

$$s_m = -\frac{\Delta f}{\Delta m} = \frac{f_0}{2m_0}. \qquad (13)$$

## 4. Device developed

Figure 4 shows the completed device. It integrates all the parts necessary for the detection of magnetoelastic signals in a single unit, such as excitation coils, detection, signal reading and control center, symmetric power supply, RF voltage gain control (modulation) and DC signal (offset). The ability to program scan settings and control the device through the GUI is an advantage that adds flexibility to the measurement and analysis process, enabling quick and easy control and recording of data. Relevant parameters in the scanning process are managed directly on its main GUI screen, such as the start frequency, end frequency and sweep step, allowing detection of ME signals with high resolution down to 0.1 Hz over a wide range of frequencies (100 Hz to 25 MHz). The induction and modulation field, resonance frequency and resonance quality factor Q are displayed on the screen, simplifying control and data recording and offering greater flexibility to customize measurements to the specific needs of each application. Table 1 summarizes the range and resolution achieved with the proposed device for ME resonance measurements.

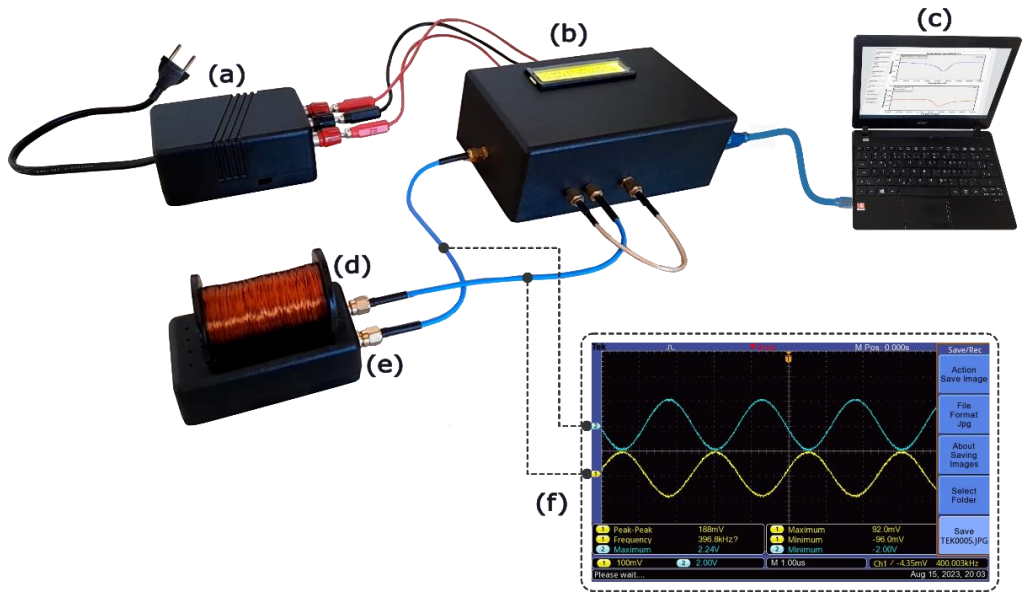

**Figure 4** - Overview of the developed prototype device. (a) symmetrical 9 V power supply, (b) electronic control unit with LCD display, (c) microcomputer, auxiliary in the detection process, (d) detection and (e) excitation coils, internal and concentric to the coil of excitation (d), where the ME sensor is positioned for reading. (f) Image of an oscilloscope screen of the output signals in blue (excitation), on a scale of 2 V per division, and input signal in yellow (detection), with a scale of 0.1 V per division.

**Table 1 -** Range and resolution of measurements with the developed device.

| Parameters | Range | Resolution |
|---|---|---|
| Frequency (excitation) | 0 a 12.5 MHz | 0.1 Hz |
| $H_{DC}$ (excitation) | -30 a 30 Oe | not measured |
| AC amplitude (excitation) | 0.2 a 3.0 V | not measured |
| Phase difference (detection) | 0 a 180° | 10 mV/° |
| Magnitude (dB) (detection) | –30 a 30 dB | 30 mV/dB |

The excitation, capture and reference signals were analyzed, which can be visualized in Fig. 5. Figure 5a shows the sinusoidal signal in blue, coming from the excitation system, with a peak-to-peak amplitude of 3.08V, obtained after going through the amplification stages. In yellow, is the reference signal, which passes through a voltage divider to have an amplitude equivalent to the signal captured by the pickup coil, presenting a peak-to-peak amplitude of 0.460V. Both signals in Fig. 5a are in phase. In Fig. 5b, the reference (yellow) and capture (blue) signals are shown. The capture signal is 90º ahead of the reference signal, as can also be seen in Fig. 4f for the excitation signal. This phase change is the result of Lenz's law, as the flux induced in the pickup coil by the excitation coil has the opposite direction, causing the signal to phase change [27]. This signal phase shift is measured by the detection system (as shown in Fig. 7b).

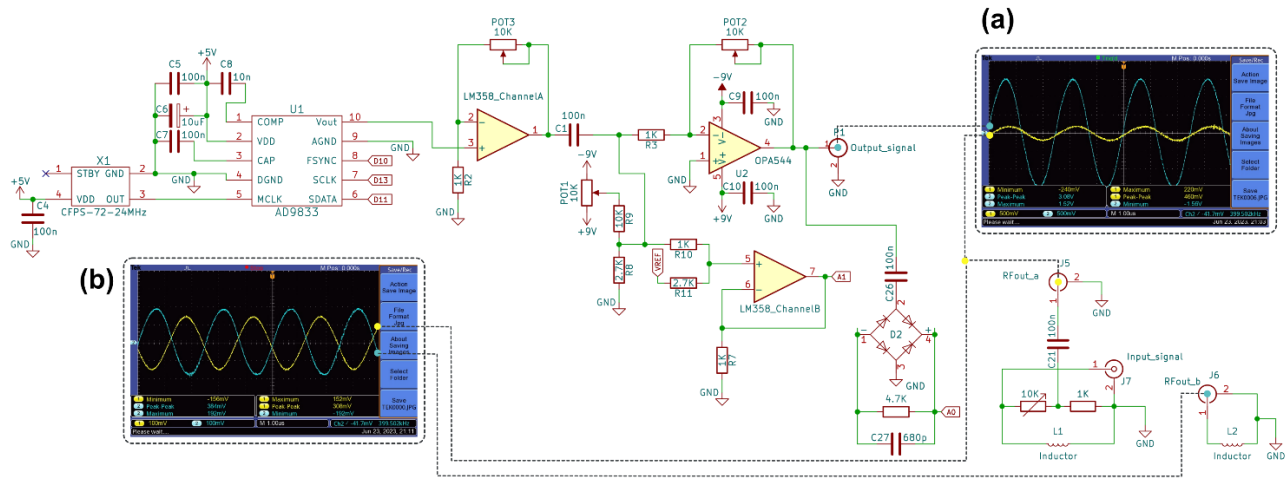

**Figure 5** - Visualization of the main signals throughout the excitation circuit. (a) Excitation (blue) and reference (yellow) signals in phase. (b) Capture (blue) and reference (yellow) signals with a phase difference of 90º, a change attributed to Lenz's Law.

The static magnetic field $H_{DC}$ along the x-direction generated by the DC bias coil is determined from the Biot-Savart law:

$$H_{DC}(x) = \frac{\mu_0 i n_{DC}}{2} \left[ \frac{\frac{L}{2} - x}{\sqrt{\left(\frac{L}{2} - x\right)^2 + R^2}} + \frac{\frac{L}{2} + x}{\sqrt{\left(\frac{L}{2} + x\right)^2 + R^2}} \right], \qquad (14)$$

where $i$ is the electric current, $n_{DC}$ is the number of turns, $L$ and $R$ are the length and the radius of the DC bias coil, respectively. Figure 6a shows the DC polarization coil. Figure 6b represents the magnetic field lines generated by it in the x-y plane. The graph in Fig. 6c shows the fit of equation (14) to experimental data for the $H_{DC}$ magnetic field along the x axis. In the region shaded in blue in Fig. 6c, the magnetic field remains constant and homogeneous throughout the ME sensor, ensuring ideal excitation conditions.

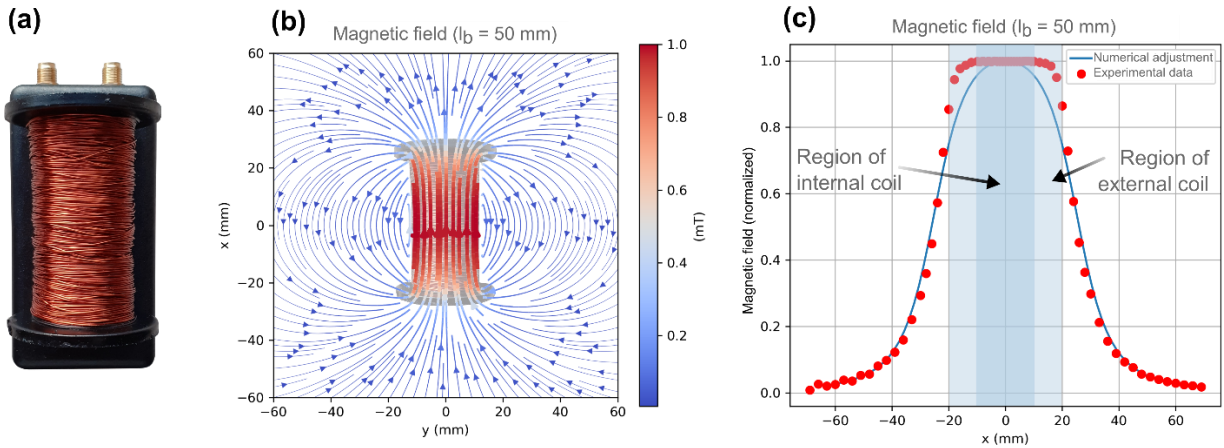

**Figure 6** - (a) Image of the DC bias coil. At its center is the pick-up coil. (b) Representation of the magnetic field lines generated by the DC bias coil in the x-y plane. (c) Blue line is the fit the experimental data to equation (14). The region comprised by the outermost blue shadow (Fig. 6c) is related to the length of the DC polarization coil, and the inner region is related to the length of the internal coil.

*4.1. Measurements obtained with the proposed device*

The results of measuring the magnetoelastic resonance frequency of the ME sensor are shown in Fig. 7, carried out with the proposed equipment in this work (Fig. 7a-b) and with the VNA (Fig. 7c-d). Both measurements were performed in the anisotropy field $H_{A\sigma} = 13.5$ Oe. In the case of different experimental configurations, when measuring with the proposed equipment, excitation occurs by the external coil and capture by the internal coil, while the VNA uses the internal coil for both excitation and capture, and the external coil is used to generate the $H_{DC}$ induction field. Therefore, although the curves and quantities analyzed present differences, the magnetoelastic resonance frequency is precisely determined in both assemblies, demonstrating the accuracy of the proposed equipment in these measurements. The resonance frequency measured by the proposed equipment is obtained at the minimum magnitude value (Fig. 7a). The resonance frequency is located at the center of the phase difference curve.

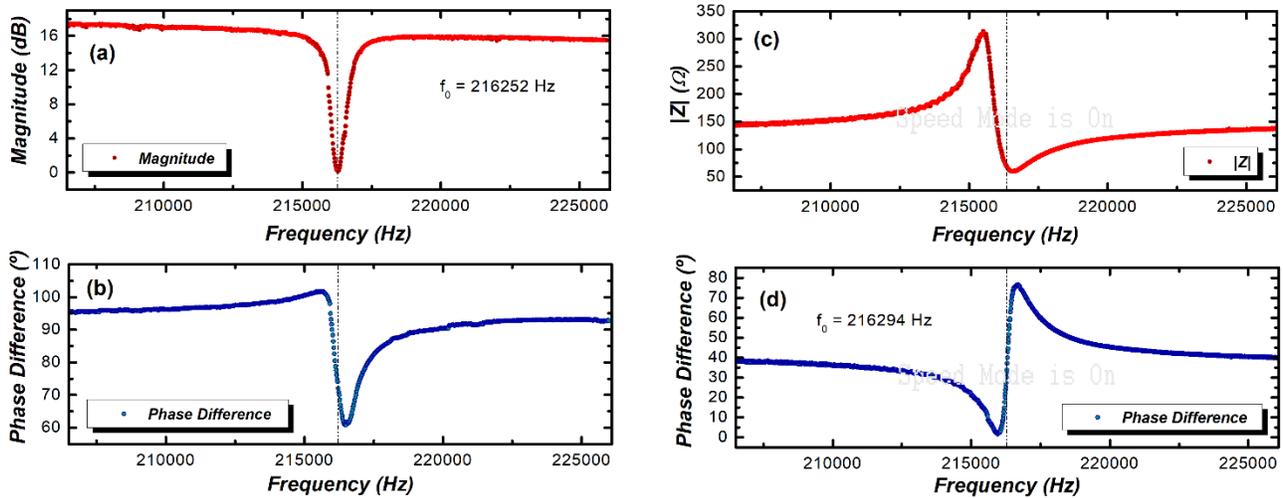

**Figure 7** – (a)-(b) Results of measuring the magnetoelastic resonance frequency of the ME sensor measuring 11 x 2 x 0.03 mm, carried out with the proposed equipment and (c)-(d) the VNA in an anisotropy field $H_{A\sigma}$ = 13.5 Oe. Both configurations show differences in the curves, due to the different experimental approaches. However, accuracy in determining the resonance frequency is obtained, highlighting the reliability of the proposed equipment.

The graphs in Fig. 8 show the results of measurements carried out with the proposed equipment of (Fig. 8a) magnitude and (Fig. 8b) phase difference measured by a sensor measured in air and immersed in water. It is verified that there is a shift in the resonance frequency as well as a drop in the oscillation amplitude measured in the sensor immersed in water when compared to the measurements carried out in air. This is due to the viscous friction that the liquid generates on the sensor surfaces, which also reflects a decrease in the resonance quality factor Q [28-29], where:

$$Q = \frac{f_r}{\Delta f} \quad \rightarrow \quad \begin{array}{l} Q_{Air} = 552.5 \\ Q_{Water} = 144.5 \end{array} \quad (15)$$

where $\Delta f$ is the width at half height on the magnitude curve. The graph in Fig. 8c represents the variation in the resonance frequency as a function of the applied magnetic field $H$. The variation observed in the resonance frequency is attributed to the "delta E" effect, indicating that the Young's modulus of the ME sensor is influenced by the magnetic field, a characteristic effect of magnetoelastic materials. From the analysis of the curve in Fig. 8c, we arrive at the value of the anisotropy field, obtained in the valley, whose value was found to be $H_{A\sigma} = 13.5$ Oe, used in this work [30].

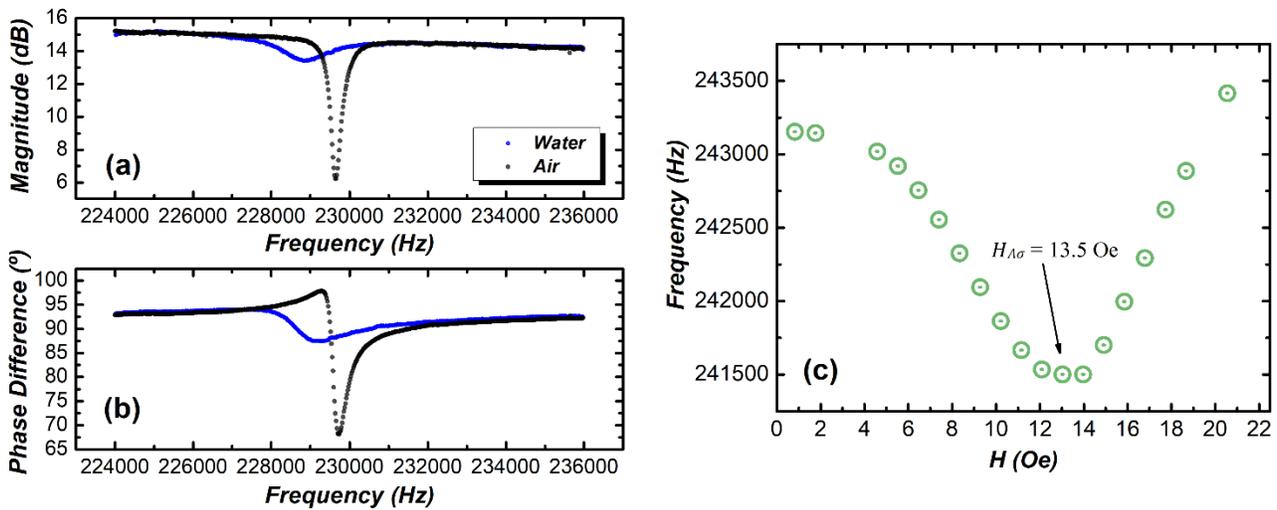

**Figure 8** - (a) Magnitude and (b) phase difference measured from a sensor in air and immersed in water. (c) Variation of the resonance frequency as a function of the applied magnetic field $H$. Such a change in the resonance frequency is attributed to the "delta E" effect, where the Young's modulus is a function of $H$, directly impacting the elastic properties of the material and, consequently, the system's resonance frequency.

The shift in the ME resonance frequency as a function of the platinum (Pt) deposit on the ME sensor was analyzed using the proposed device. The graph in Fig. 9a shows the presence of the shift in the ME resonance frequency for different thicknesses of Pt deposited on the sensor, always moving to the left. A small downward shift at the base of the curve signals (50 - 250 nm) in relation to the signal without deposition (0 nm) is due to a possible influence of the deposited film on the magnetic susceptivity of the medium, which is measured by the $log\left(\frac{V_{INPA}}{V_{INPB}}\right)$ expressed in equation (1). A small drop in signal amplitude for depositions (50 - 250 nm) in relation to the signal without deposition (0 nm) is also observed, a possible contribution from the tension formed between the substrate and the deposited film, which may lead to this decrease in amplitude.

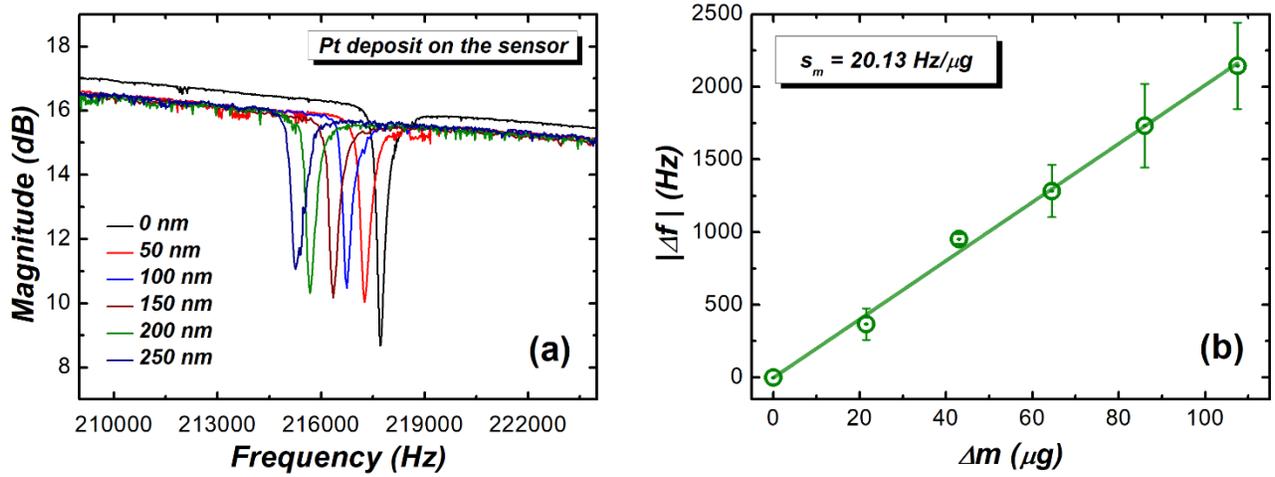

**Figure 9** - (a) Shift of the ME resonance frequency as a function of the thickness of the deposited film. (b) Absolute shift of the ME resonance frequency for different masses associated with films of different thicknesses deposited.

The graph in Fig. 9b shows the absolute displacement of the ME resonance frequency for different masses associated with films of different thicknesses deposited, considering the sensor area $A = 0.00005\ m^2$ and the density of platinum $\rho_{Pt} = 21500\ \frac{Kg}{m^3}$. In this way, we can arrive at the sensitivity factor E described by equation (13), whose value obtained was $s_m = 20.13\ \frac{Hz}{\mu g}$. The sensitivity $s_m$ is directly related to the mass of the bare sensor, as described in equation (13), that is, the smaller the sensor, the more sensitive it is. Close values for $s_m$, considering the size of the sensor used. Saiz et al. [1] obtained values of $s_m = 47.4 \frac{Hz}{\mu g}$ for a sensor using the same Metglas 2826 MB alloy with dimension 1 mm × 10 mm × 28 μm and $s_m = 12.3 \frac{Hz}{\mu g}$ for a sensor dimension 1.5 mm × 15 mm × 28 μm, for different deposited gold (Au) thicknesses. Ersöz et al. [31] using the same Metglas 2826MB alloy with dimensions of approximately 55 mm × 6.5 mm × 30 μm, found lower sensitivity values, around $s_m = 1 \frac{Hz}{\mu g}$.

## 5. Patent

This device and the GUI software are registered with the Brazilian National Institute of Intellectual Property (INPI, Ministry of Economy, Brazil) under registration numbers BR102024008915-4 and BR512024000754-7, respectively.

## 6. Conclusions

The study presented the development of an integrated device based on a gain/phase detector and the Arduino platform for measuring magnetoelastic resonance, standing out for its simplicity, low cost, and efficiency. Utilizing the ATmega328P microcontroller of the Arduino coupled with the AD8302 and an intuitive graphical interface developed in Python, the device facilitates the measurement of magnetoelastic signals in sensors, making the technology accessible to various fields of research. The validation of the device was carried out through measurements of mass deposits on ME sensors, confirming its accuracy and functionality. This advancement provides a practical and portable solution for detecting magnetoelastic resonances, promoting the use of these technologies in field environments and other scientific applications.

**Acknowledgements**

This research is supported by Conselho Nacional de Desenvolvimento Científico e Tecnológico (CNPq), Coordenação de Aperfeiçoamento de Pessoal de Nível Superior (CAPES), Financiadora de Estudos e